\documentstyle[11pt,epsf]{article}
\newtheorem{theorem}{Theorem}

\newcommand {\dfn} {\stackrel{\Delta} {=}}
\newcommand {\exe} {\stackrel{\cdot} {=}}
\newcommand {\gexe} {\stackrel{\cdot} {\ge}}
\newcommand {\lexe} {\stackrel{\cdot} {\le}}

\newcommand {\reals} {{\rm I\!R}}

\newcommand {\bs} {\mbox{\boldmath $s$}}
\newcommand {\bt} {\mbox{\boldmath $t$}}

\newcommand {\bsigma} {\mbox{\boldmath $\sigma$}}
\newcommand {\bx} {\mbox{\boldmath $x$}}
\newcommand {\by} {\mbox{\boldmath $y$}}

\newcommand {\bE} {\mbox{\boldmath $E$}}

\newcommand {\bX} {\mbox{\boldmath $X$}}
\newcommand {\bY} {\mbox{\boldmath $Y$}}

\newcommand{\calA}{{\cal A}}
\newcommand{\calB}{{\cal B}}
\newcommand{\calC}{{\cal C}}
\newcommand{\calD}{{\cal D}}

\newcommand{\calI}{{\cal I}}
\newcommand{\calJ}{{\cal J}}

\newcommand{\calM}{{\cal M}}

\newcommand{\calS}{{\cal S}}
\newcommand{\calT}{{\cal T}}
\newcommand{\calU}{{\cal U}}

\newcommand{\calX}{{\cal X}}
\newcommand{\calY}{{\cal Y}}

\begin{document}
\thispagestyle{empty}
\title{
Universal Decoding for Arbitrary Channels 
Relative to a Given Class of Decoding Metrics
}
\author{Neri Merhav
}
\date{}
\maketitle

\begin{center}
Department of Electrical Engineering \\
Technion - Israel Institute of Technology \\
Technion City, Haifa 32000, ISRAEL \\
E--mail: {\tt merhav@ee.technion.ac.il}\\
\end{center}
\vspace{1.5\baselineskip}
\setlength{\baselineskip}{1.5\baselineskip}

\begin{abstract}
We consider the problem of universal decoding for arbitrary unknown channels 
in the random coding regime. For
a given random coding distribution and a given class of metric decoders,
we propose a generic universal decoder whose average error
probability is, within a sub--exponential multiplicative factor, no larger
than that of the best decoder within this class of decoders. Since the optimum,
maximum likelihood (ML) decoder of the underlying channel is not necessarily assumed to
belong to the given class of decoders, this setting suggests a common
generalized framework for: (i) mismatched decoding, (ii) universal
decoding for a given family of channels, and (iii) universal
coding and decoding for deterministic channels using the individual--sequence
approach. The proof of our universality
result is fairly simple, and it is demonstrated how some earlier results on
universal decoding are obtained as special cases. We also demonstrate how our
method extends to more complicated scenarios, like incorporation of noiseless
feedback, and the multiple access channel.
\end{abstract}


\newpage

\section{Introduction}

In many situations practically encountered in coded communication
systems, channel uncertainty and variability 
preclude the implementation of
the optimum maximum likelihood
(ML) decoder, and so, universal decoders, independent of the unknown channel,
are sought. 

The topic of universal coding and decoding under channel uncertainty has received
very much attention in the last four decades. In \cite{Goppa75}, Goppa
offered the {\it maximum mutual
information} (MMI) decoder, which decides in favor of the codeword having the
maximum empirical mutual information with the channel output sequence. Goppa showed that for
discrete memoryless channels (DMC's), MMI
decoding achieves capacity. Csisz\'ar and K\"orner \cite{CK81} have also
studied the problem of universal
decoding for DMC's with finite input and output alphabets. They showed that the
random coding error exponent of the MMI decoder, associated with a uniform random
coding distribution over a certain type class, achieves the optimum random
coding error exponent.
Csisz\'ar \cite{Csiszar82} proved that
for any modulo--additive DMC and the uniform
random coding distribution over linear codes, the optimum random coding error exponent is
universally achieved by a decoder that
minimizes the empirical entropy of the difference between the output sequence
and the input sequence. In \cite{Merhav93} an analogous result was derived for
a certain parametric class of memoryless Gaussian channels with an unknown interference
signal. 

In the realm of channels with memory,
Ziv \cite{Ziv85} explored the universal decoding problem for unknown finite--state channels
with finite input
and output alphabets, for which the next channel state is a deterministic
unknown function (a.k.a.\ the next--state function) 
of the channel current state and current inputs and outputs. For codes governed by
uniform random coding over a given set, he proved that a decoder based on the
Lempel--Ziv algorithm asymptotically achieves the error exponent associated with ML
decoding. In \cite{LZ98}, Lapidoth and Ziv proved that the latter decoder continues to be universally
asymptotically optimum in the random coding error exponent sense even for a wider class
of finite--state channels, namely, those with stochastic, rather than
deterministic, next--state functions.
In \cite{FL98}, Feder and Lapidoth furnished sufficient conditions for
families of channels with memory
to have universal decoders that asymptotically achieve the random coding error
exponent associated with ML decoding. In \cite{FM02}, Feder and Merhav 
proposed a competitive minimax criterion, in an effort to develop a
more general systematic approach to the problem of universal decoding. According to
this approach, an optimum decoder is sought in the quest for
minimizing (over all decision rules) the maximum (over
all channels in the family) ratio between the error probability associated with a given
channel and a given decision rule, and the error probability of the ML decoder for that
channel, possibly raised some power less than unity.

More recently, interesting attempts (see, e.g., \cite{LF12a}, \cite{LF12b},
\cite{MW12}, \cite{SF05})
were made to devise coding and decoding strategies
that avoid any probabilistic
assumptions concerning the operation of the channel. This is in the spirit of the
individual--sequence approach in information theory, 
that was originally developed in universal source coding
\cite{ZL78} and later on further exercised in other problem areas. 
In \cite{LF12a}, the notion of {\it empirical rate functions} has been
established and investigated (with and without feedback) 
for a given input distribution and for given posterior probability function
(or a family of such functions) of the channel input sequence given the output
sequence. In
\cite{MW12}, capacity--achieving (or ``porosity--achieving'', in the
terminology of \cite{MW12})
universal encoders and decoders, namely, encoder--decoder pairs with coding rates as high as the best
finite--state encoder and decoder,
were devised for 
modulo additive channels with deterministic noise sequences and noiseless
feedback. This feedback is necessary to let the encoder adapt to the channel,
which otherwise does not access the channel output and thus 
cannot learn (either implicitly or explicitly) the
characteristics of the channel.

In this paper, we take a somewhat different approach.
We consider the problem of universal decoding for {\it arbitrary} unknown channels
in the random coding regime. For
a given random coding distribution and a given class of metric decoders,
we propose a generic universal decoder whose average error
probability is, within a sub--exponential multiplicative factor, no larger
than that of the best decoder in this class of decoders. Since the optimum,
ML decoder of the underlying channel is not necessarily
assumed to belong to the given class of decoders, this setting is suitable as a common
ground for: 
\begin{enumerate}
\item Mismatched decoding (see,
e.g., \cite{CN95}, \cite{Hui83}, \cite{MKLS94}) --
when the reference class of decoders is a singleton and the ML decoder for the underlying
channel is different from the unique decoder in this singleton. 
\item Universal decoding for a given family of channels (as in papers cited in
the second and third paragraphs above) -- when the ML decoder for the underlying channel
belongs to the given class of decoders.  
\item Universal
coding and decoding for deterministic channels using the individual--sequence
approach (as in \cite{LF12a}, \cite{LF12b}, \cite{MW12}, 
\cite{SF05}) -- when the underlying channel is deterministic
and the universality is relative to a given class of coding/decoding strategies.
\end{enumerate}
The proof of our universality
result is fairly simple and general, and it is demonstrated how some earlier
mentioned results on
universal decoding are obtained as special cases. It is based on very simple
upper and lower bounds on the probabilities of the pairwise error events, as
well as on a lower bound due to Shulman \cite[Lemma A.2]{Shulman03} 
on the probability of the union of pairwise
independent events, which coincides with the union bound up to a factor of
1/2. 

Finally, we demonstrate how our
method extends to more complicated scenarios. The first extension
corresponds to random coding distributions that allow to incorporate noiseless feedback.
This extension is fairly straightforward, but its main importance is in allowing
adaptation of the random coding distribution 
to the channel statistical characteristics. The second extension is
to the problem of universal decoding for multiple access channels (MAC's)
with respect to a given class of decoding metrics. 
This extension is not trivial since the
universal decoding metric has to confront three different types of error events (in the
case of a MAC with two senders). In particular, it turns out that
the resulting universal decoding metric 
is surprisingly different from those of earlier works on universal decoding for the
MAC \cite{LH96}, \cite[Section VIII]{FL98}, \cite{PW85}, mostly because the
problem setting here is different from those of these earlier works (in the
sense that the
universality here is relative to a given class of decoders while the underlying
channel is arbitrary, and not relative to a
given class of channels).

The outline of the paper is as follows. In Section 2, we establish notation
conventions and we formalize the problem setting. Section 3 contains our main
result and its proof, as well as a discussion and examples. Section 4 suggests
guidelines for approximating the universal decoding metric in situations
where it is hard to compute, and thereby shows how Ziv's decoding metric
\cite{Ziv85} falls within our framework. Finally, in Section 5, we provide
extensions to the case where feedback
is available and the case of the MAC.

\section{Notation Conventions and Problem Formulation}

\subsection{Notation Conventions}

Throughout this paper, scalar random variables (RV's) are denoted by
capital letters, their sample
values are denoted by the respective lower case letters, and their alphabets
are denoted by the
respective calligraphic letters. A similar convention applies to random
vectors of dimension $n$ and
their sample values, which will be denoted with same symbols in the bold face
font. The set of all
$n$--vectors with components taking values in a certain alphabet, will
be denoted as the same
alphabet superscripted by $n$. Sources and channels will be denoted generically
by the letter $P$ or $Q$. For example, the channel input probability distribution
function will be denoted by $Q(\bx)$, $\bx\in\calX^n$, and the conditional
probability distribution of the channel output vector $\by\in\calY^n$ given the input
vector $\bx\in\calX^n$, will be denoted by $P(\by|\bx)$.
Information theoretic quantities like entropies and conditional entropies, will
be denoted following
the standard conventions of the information theory literature, e.g., $H(\bX)$, $H(\bX|\bY)$, etc. 
The expectation operator will be denoted by $\bE\{\cdot\}$ and
the cardinality of a
finite set $\calA$ will be denoted by $|\calA|$.

For a given sequence $\bx\in\calX^n$, $\calX$ being a finite alphabet,
$\hat{P}_{\bx}$
denotes the empirical distribution on $\calX$ extracted 
from $\bx$, in other words,
$\hat{P}_{\bx}$ is the vector $\{\hat{P}_{\bx}(x),~x\in\calX\}$, where
$\hat{P}_{\bx}(x)$
is the relative frequency of the letter $x$ in the vector $\bx$.
The type class of $\bx$, denoted $T_{\bx}$, is the set of all sequences
$\bx'\in\calX^n$ with $\hat{P}_{\bx'}=\hat{P}_{\bx}$. Similarly, for a pair of
sequences $(\bx,\by)\in\calX^n\times\calY^n$, the empirical distribution
$\hat{P}_{\bx\by}$ is the matrix of relative frequencies
$\{\hat{P}_{\bx\by}(x,y),~x\in\calX,~y\in\calY\}$ and the type class
$T_{\bx\by}$ is the set of pairs $(\bx',\by')\in\calX^n\times\calY^n$ with
$\hat{P}_{\bx'\by'}=\hat{P}_{\bx\by}$. For a given $\by$, $T_{\bx|\by}$
denotes the conditional type class of $\bx$ given $\by$, which is the set of
vectors $\{\bx'\}$ such that $(\bx',\by)\in T_{\bx\by}$.
Information measures induced by empirical distributions, i.e.,
empirical information measures, will be denoted with a
hat and a subscript that indicates the sequence(s) from which they are
induced. For example, $\hat{H}_{\bx}(X)$ is the empirical entropy extracted
from $\bx\in\calX^n$, namely, the entropy of a random variable $X$ whose
distribution is $\hat{P}_{\bx}$. Similarly, $\hat{H}_{\bx\by}(X|Y)$ and
$\hat{I}_{\bx\by}(X;Y)$ are, respectively, the 
empirical conditional entropy of $X$ given $Y$, and the empirical mutual
information between $X$ and $Y$, extracted from $(\bx,\by)$, and so on.

For two sequences of positive numbers, $\{a_n\}$ and $\{b_n\}$, the notation
$a_n\exe b_n$ means that
$\frac{1}{n}\log\frac{a_n}{b_n}\to 0$
as $n\to\infty$. Similarly, $a_n\lexe b_n$ means that
$\limsup_{n\to\infty}\frac{1}{n}\log\frac{a_n}{b_n}\le 0$, and so on.
The functions $\log(\cdot)$ and $\exp(\cdot)$, throughout this paper, 
will be defined to the base 2, unless
otherwise indicated. The operation $[\cdot]_+$ will mean positive clipping,
that is $[x]_+=\max\{0,x\}$.

\subsection{Problem Formulation}

Consider a random selection of a codebook 
$\calC=\{\bx_1,\ldots,\bx_M\}\subseteq\calX^n$, where $M=2^{nR}$, $R$ being
the coding rate in bits per channel use.  
The marginal probability distribution function of each
codeword $\bx_i$ is denoted by $Q(\bx_i)$. It will be assumed that the various codewords
are pairwise independent.\footnote{Full independence of all codewords is
allowed, but not enforced. This permits our setting to include, among other
things, ensembles of linear codes,
which are well known to admit pairwise independence, but not stronger notions
of independence.} Let $P(\by|\bx)$ be the conditional probability distribution
of the channel output vector $\by\in\calY^n$ given the channel input vector
$\bx\in\calX^n$. We make no assumptions at all concerning the
channel.\footnote{We even allow a deterministic channel, which puts all its
probabilistic mass on one vector $\by$ which is given by a deterministic
function of $\bx$.}
We will assume, throughout most of this paper, that both the
channel input alphabet $\calX$ and
the channel output alphabet $\calY$ are finite sets. Finally, we define a
class of {\it decoding metrics}, as a class of real functions,
$\calM=\{m_\theta(\bx,\by),~\theta\in\Theta,~\bx\in\calX^n,~\by\in\calY^n\}$,
where $\Theta$ is an index set, which may be either
finite, countably infinite, or uncountably infinite.\footnote{For example, in the
uncountably infinite case, $\theta$ may designate a parameter
and $\{m_\theta(\bx,\by),~\theta\in\Theta\}$ may be a smooth parametric
family.} The decoder associated with the decoding metric $m_\theta$, which
will be denoted by $\calD_\theta$, decides in
favor of the message $i\in\{1,\ldots,M\}$ which maximizes
$m_\theta(\bx_i,\by)$ for the given received 
channel output vector $\by$, that is
\begin{equation}
\calD_\theta:~~~~\hat{i}=\mbox{argmax}_{1\le i\le M}m_\theta(\bx_i,\by).
\end{equation}
The message $i$ is assumed to be uniformly distributed in the set $\{1,2,\ldots,M\}$.
It should be emphasized that the optimum, ML decoding metric for the
underlying channel
$P(\by|\bx)$, may not necessarily belong to the given class of decoding
metrics $\calM$. In other words, this is a problem of universal decoding
with possible mismatch.

The average error probability $\bar{P}_{e,\theta}(R,n)$, associated with the
decoder $\calD_\theta$, is defined as
\begin{equation}
\bar{P}_{e,\theta}(R,n)\dfn\frac{1}{M}\sum_{i=1}^M\mbox{Pr}\bigcup_{j\ne
i}\left\{m_\theta(\bX_j,\bY)\ge
m_\theta(\bX_i,\bY)\bigg|\bX_i~\mbox{sent}\right\},
\end{equation}
where $\mbox{Pr}\{\cdot\}$ designates the probability measure pertaining to the
randomness of the codebook $\calC$ as well as that of the channel output given its
input.

While the decoder $\calD_\theta$, that minimizes
$\bar{P}_{e,\theta}(R,n)$ within the class, depends, in general, on the unknown
underlying channel $P(\by|\bx)$, our goal is to devise a universal decoder
$\calU$, with a decoding metric $U(\bx,\by)$, independent of
the underlying channel $P(\by|\bx)$, whose average error probability would be
essentially as small as $\min_\theta \bar{P}_{e,\theta}(R,n)$, whatever the
underlying channel may be. 
By ``essentially as small'', we mean that the
average error probability associated with the universal decoder,
\begin{equation}
\bar{P}_{e,u}(R,n)\dfn
\frac{1}{M}\sum_{i=1}^M\mbox{Pr}\bigcup_{j\ne
i}\left\{U(\bX_j,\bY)\ge
U(\bX_i,\bY)\bigg|\bX_i~\mbox{sent}\right\},
\end{equation}
would not
exceed $\min_\theta \bar{P}_{e,\theta}(R,n)$ by more than a multiplicative factor that grows
sub--exponentially with $n$. This means that whenever $\min_\theta
\bar{P}_{e,\theta}(R,n)$ decays exponentially with $n$, then so does
$\bar{P}_{e,u}(R,n)$, and at an exponential rate at least as fast.
Another (essentially equivalent) legitimate goal is that
$\bar{P}_{e,u}(R,n)$ would not be larger than $\min_\theta
\bar{P}_{e,\theta}(R+\Delta_n,n)$, where $\Delta_n\to 0$ as $n\to \infty$.
In the next section, we shall see that both goals are met by a conceptually simple
universal decoding metric $U(\bx,\by)$, which depends solely on $Q$ and on the reference class $\calM$
of competing decoding metrics.

\section{Main Result}

Consider the given random coding distribution $Q$ and the given family of
decoding metrics $\calM=\{m_\theta(\bx,\by)~,\theta\in\Theta\}$,
as defined earlier.
Let us define
\begin{equation}
\calT(\bx|\by)\dfn\left\{\bx^\prime:~~\forall
\theta\in\Theta~~m_\theta(\bx^\prime,\by)=
m_\theta(\bx,\by)\right\}.
\end{equation}
Our universal decoding metric is defined as
\begin{equation}
\label{universalmetric}
U(\bx,\by)\dfn-\frac{1}{n}\log Q[\calT(\bx|\by)].
\end{equation}
Note that when $\calX$ is a discrete alphabet, 
$\{\calT(\bx|\by)\}$ are equivalence classes for every
$\by\in\calY^n$, and so the space $\calX^n$ can be partitioned into a disjoint
union of them. Let $K_n(\by)$ denote the number of equivalence classes
$\{\calT(\bx|\by)\}$ for a given $\by$. Also define 
\begin{equation}
K_n\dfn\max_{\by\in\calY^n}K_n(\by)
\end{equation}
and
\begin{equation}
\Delta_n\dfn \frac{\log K_n}{n}.
\end{equation}
Our main result is the following theorem:
\begin{theorem}
Under the assumptions of Section 2,
the universal decoding metric defined in eq.\ (\ref{universalmetric})
satisfies:
\begin{equation}
\bar{P}_{e,u}(R,n)\le 2\cdot 2^{n\Delta_n}\cdot\min_{\theta\in\Theta}
\bar{P}_{e,\theta}(R,n)
\end{equation}
and
\begin{equation}
\bar{P}_{e,u}(R,n)\le 2\cdot\min_{\theta\in\Theta}
\bar{P}_{e,\theta}(R+\Delta_n,n).
\end{equation}
\end{theorem}

\noindent
{\it Discussion.} The theorem is, of course, meaningful when $\Delta_n\to 0$ as $n\to\infty$,
which means that the number of various equivalence classes
$\{\calT(\bx|\by)\}$ grows sub-exponentially as a function of $n$, uniformly in
$\by$. As mentioned earlier, in this case, whenever $\min_{\theta\in\Theta}\bar{P}_{e,\theta}(R,n)$ decays
exponentially with $n$, then $\bar{P}_{e,u}(R,n)$ decays exponentially as well,
and at least as fast. Consequently, the maximum information rate pertaining to
the universal decoder is at least as large as that of the best decoder
$\calD_\theta$ in the given class. We therefore learn from Theorem 1 that a
sufficient condition for the existence of a universal decoder is
$\lim_{n\to\infty}\Delta_n=0$. Whether this is also a necessary condition,
remains an open question at this point. Necessary and sufficient conditions for
universality in the ordinary setting have been furnished in \cite{FL98} and
\cite{FM02}. 

Intuitively, the behavior of $\Delta_n$ for large $n$ is a measure of
the richness of the class of decoding metrics. The larger is the index set 
$\Theta$, the smaller are the equivalence classes $\{\calT(\bx|\by)\}$, and
then their total number $K_n(\by)$ 
becomes larger, and so does $\Delta_n$. Universality is enabled, using this
method, as long as the set $\Theta$ is not too rich, so that $\Delta_n$ still
vanishes as $n$ grows without bound.

When $Q$ is invariant within $\calT(\bx|\by)$ (i.e., $\bx'\in\calT(\bx|\by)$
implies $Q(\bx')=Q(\bx)$), we have
\begin{eqnarray}
U(\bx,\by)&=&-\frac{1}{n}\log Q[\calT(\bx|\by)]\nonumber\\
&=&-\frac{1}{n}\log [Q(\bx)\cdot|\calT(\bx|\by)|]\nonumber\\
&=&-\frac{1}{n}[\log Q(\bx)+\log|\calT(\bx|\by)|].
\end{eqnarray}
The choice of a distribution $Q$ that is invariant within $T(\bx|\by)$ is
convenient, because in most cases it is easier to evaluate the log--cardinality of
$\calT(\bx|\by)$ (or its log--volume, in the continuous case) than to evaluate
its probability under a general probability measure $Q$.

Before we turn to the proof of Theorem 1, it would be instructive to consider
two simple examples. In both of them (as well as in other examples in the
sequel) $Q$ is invariant within $\calT(\bx|\by)$.

\noindent
{\it Example 1.}
Let $Q$ be the uniform distribution across a single type class,
$T_{\bx}$, and let $\calM$ be the class of additive decoding metrics
\begin{equation}
m_\theta(\bx,\by)=\sum_{i=1}^n\theta(x_i,y_i),
\end{equation}
where $\{\theta(x,y),~x\in\calX,~y\in\calY\}$ are arbitrary real--valued
matrices. In this case, $\calT(\bx|\by)=T_{\bx|\by}$, the conditional type class of $\bx$
given $\by$. Since the number of distinct conditional type classes is
polynomial in $n$, then $\Delta_n$ is proportional to $(\log n)/n$.
In this case, we have
\begin{eqnarray}
U(\bx,\by)&=&-\frac{1}{n}\log Q[T_{\bx|\by}]\\
&=&-\frac{1}{n}\log [Q(\bx)\cdot|T_{\bx|\by}|]\\
&=&\hat{H}_{\bx}(X)-\hat{H}_{\bx\by}(X|Y)+o(n)\\
&=&\hat{I}_{\bx\by}(X;Y)+o(n).
\end{eqnarray}
and so, the proposed universal decoder essentially\footnote{The $o(n)$
term can be omitted with affecting the asymptotic performance.} coincides with the MMI decoder.
However, since $\calC$ is a constant composition code, under this particular choice of $Q$,
$\hat{H}_{\bx_i}(X)$ is the same for all $i$, and so,
this decoder is equivalent to the decoder that selects the codeword that minimizes the empirical 
conditional entropy of $X$ given $Y$, namely, $\min_i\hat{H}_{\bx_i\by}(X|Y)$.
If, on the other hand, $Q$ is an i.i.d.\ probability distribution function,
namely, $Q(\bx)=\prod_{i=1}^nQ(x_i)$, then the universal decoding metric
becomes
\begin{equation}
U(\bx,\by)=\hat{I}_{\bx\by}(X;Y)+D(\hat{P}_{\bx}\|Q)+o(n),
\end{equation}
where $D(\hat{P}_{\bx}\|Q)$ is the Kullback--Leibler divergence between
$\hat{P}_{\bx}$ and $Q$. 

For certain classes of channels (e.g.,
arbitrarily varying channels), it is not difficult
to derive single--letter formulas for the maximum achievable information rates
in the random coding regime, that is, the
supremum of $R$ such that $\bar{P}_{e,u}(R,n)\to 0$
as $n\to\infty$. The main tool for this purpose is the method of
types. This concludes Example 1. $\Box$

\noindent
{\it Example 2.} Let $\calX=\calY=\reals$ and let 
\begin{equation}
Q(\bx)=\frac{e^{-\sum_{i=1}^nx_i^2/(2\sigma^2)}}{(2\pi\sigma^2)^{n/2}}.
\end{equation}
Let $\theta=(\theta_1,\theta_2)\in\reals^2$ and
$\calM$ be the class of decoding metrics of the form\footnote{This class of
decoders is clearly motivated by the family of channels
$y_t=ax_t+z_t$, where $a$ is an unknown parameter and $z_t$ is an i.i.d.\
Gaussian process, independent of $x_t$.}
\begin{equation}
m_\theta(\bx,\by)=\theta_1\sum_{i=1}^nx_iy_i+\theta_2\sum_{i=1}^nx_i^2.
\end{equation}
In principle, $\calT(\bx|\by)$ is the set of all $\{\bx'\}$ with the same
empirical power
and the same empirical correlation with $\by$, 
as those of $\bx$. However,
since in this example the sequences $\bx$ and $\by$ have continuous--valued
components, some tolerance must be allowed in the empirical
correlation $C(\bx,\by)=\frac{1}{n}\sum_{i=1}^nx_iy_i$ 
and empirical power, $S(\bx)=\frac{1}{n}\sum_{i=1}^nx_i^2$, 
for $\calT(\bx|\by)$ to have positive probability (and positive volume),
and so, $\calT(\bx|\by)$ should be redefined as the
set of sequences $\bx'$, where $C(\bx',\by)$ and $S(\bx')$ are within
$\epsilon$ ($\epsilon> 0$, but small) close to $C(\bx,\by)$ and $S(\bx)$,
respectively. Using the methods developed in \cite{Merhav93},\footnote{The
details are conceptually simple but technically tedious. The interested
reader is referred to \cite{Merhav93} for a rigorous treatment.}
it is not difficult to show that, after omitting some additive constants
(which do not affect the decision rule),
we have in this case
\begin{equation}
U(\bx,\by)=\frac{S(\bx)}{2\sigma^2}-\frac{1}{2}\ln[S(\bx)(1-\rho_{\bx\by}^2)],
\end{equation}
where $\rho_{\bx\by}=C(\bx,\by)/\sqrt{S(\bx)S(\by)}$ is the empirical
correlation coefficient between $\bx$ and $\by$, and where
we have used natural logarithms instead of base 2 logarithms for obvious
reasons. The first term stems from
$-\frac{1}{n}\ln Q(\bx)$ and the second term comes from the negative log--volume of
$\calT(\bx|\by)$. 
This concludes Example 2. $\Box$

\noindent
{\it Proof of Theorem 1.}
The pairwise average error probability,
associated with $m_\theta$ is lower bounded by
\begin{eqnarray}
\label{lower}
\bar{\Pi}_{e,\theta}(\bx,\by)&\dfn&\sum_{\{\bx^\prime:~m_\theta(\bx^\prime,\by)\ge
m_\theta(\bx,\by)\}}Q(\bx^\prime)\\
&\geq&\sum_{\bx^\prime\in \calT(\bx|\by)}Q(\bx^\prime)\\
&=&Q[\calT(\bx|\by)]\\
&=&\exp[-nU(\bx,\by)].
\end{eqnarray}
On the other hand, the pairwise error probability associated with the decoding
metric $U$ is upper bounded by
\begin{eqnarray}
\label{upper}
\bar{\Pi}_{e,u}(\bx,\by)&\dfn&\sum_{\{\bx^\prime:~U(\bx^\prime,\by)\ge
U(\bx,\by)\}}Q(\bx^\prime)\\
&=&\sum_{\{\calT(\bx^\prime|\by):~U(\bx^\prime,\by)\ge U(\bx,\by)\}}
\sum_{\tilde{\bx}\in\calT(\bx^\prime|\by)}Q(\tilde{\bx})\\
&=&\sum_{\{\calT(\bx^\prime|\by):~U(\bx^\prime,\by)\ge U(\bx,\by)\}}
Q[\calT(\bx^\prime|\by)]\\
&=&\sum_{\{\calT(\bx^\prime|\by):~U(\bx^\prime,\by)\ge U(\bx,\by)\}}
\exp[-nU(\bx^\prime,\by)]\\
&\le&\sum_{\{\calT(\bx^\prime|\by):~U(\bx^\prime,\by)\ge U(\bx,\by)\}}
\exp[-nU(\bx,\by)]\\
&\le&\sum_{\{\calT(\bx^\prime|\by)\}}
\exp[-nU(\bx,\by)]\\
&\leq&2^{n\Delta_n}\exp[-nU(\bx,\by)]\\
&=&\exp\{-n[U(\bx,\by)-\Delta_n]\},
\end{eqnarray}
where in the second equality we have used the fact that $U(\bx,\by)$ depends
on $\bx$ and $\by$ only via $\calT(\bx|\by)$ and the last
inequality follows from the fact that the number of different equivalence classes
$\{\calT(\bx'|\by)\}$ is upper bounded by $K_n=2^{n\Delta_n}$ by definition.
Now, as is well known, given $\bx$ and $\by$, the average probability of error can be upper bounded in
terms of the average pairwise error probability by the expectation of the
union bound, clipped to unity, that is
\begin{equation}
\label{ub}
\bar{P}_{e,u}(R,n)\le
\bE\left[\min\left\{1,2^{nR}\bar{\Pi}_{e,u}(\bX,\bY)\right\}\right]\le
\bE\left[\min\left\{1,2^{nR}\exp(-n[U(\bX,\bY)-\Delta_n])\right\}\right],
\end{equation}
where the expectation is w.r.t.\ the randomness of $\bX$ and $\bY$, whose
joint distribution is given by $Q(\bx)P(\by|\bx)$.

Next, we need a lower bound on $\bar{P}_{e,\theta}(R,n)$ in terms of
$\Pi_{e,\theta}(\bx,\by)$. To this end, we invoke the following lower bound on the
probability of the union of pairwise independent events $\calA_1,\ldots,\calA_M$, proved by
Shulman \cite[p.\ 109, Lemma A.2]{Shulman03}\footnote{A similar result was
proved independently in \cite[Lemma 1]{SBM07} for fully independent events
with equal probabilities.}
\begin{equation}
\mbox{Pr}\left\{\bigcup_{i=1}^M\calA_i\right\}\ge
\frac{1}{2}\cdot\min\left\{1,\sum_{i=1}^M\mbox{Pr}(\calA_i)\right\}.
\end{equation}
In our case, for a given $\bx_i=\bx$ and $\by$, the events
$\{m_\theta(\bX_j,\by)\ge m_\theta(\bx,\by)\}_{j\ne i}$ are pairwise independent
since we have assumed that the various codewords are pairwise independent.
Thus, after taking the expectation w.r.t.\ the joint distribution of
$(\bX,\bY)$, we have
\begin{equation}
\label{lb}
\bar{P}_{e,\theta}(R,n)\ge\frac{1}{2}\cdot
\bE\left[\min\left\{1,2^{nR}\bar{\Pi}_{e,\theta}(\bX,\bY)\right\}\right]
\ge\frac{1}{2}\cdot\bE\left[\min\left\{1,2^{nR}\exp(-nU(\bX,\bY)\right\}\right].
\end{equation}
Comparing now the right--most side of eq.\ (\ref{ub}) with that of eq.\
(\ref{lb}), we readily see that $\bar{P}_{e,u}(R,n)$ is upper bounded both by
$2\bar{P}_{e,\theta}(R+\Delta_n,n)$ and by
$2\cdot 2^{n\Delta_n}\bar{P}_{e,\theta}(R,n)$. The first upper bound is obtained
by combining $\Delta_n$ and $R$ in (\ref{ub}) and the
second upper bound is
obtained similarly, by upper bounding the unity term (in $\min\{1,
2^{n[R+\Delta_n]}\bar{\Pi}_{e,\theta}(\bX,\bY)\}$) 
by $2^{n\Delta_n}$, which then becomes
a constant multiplicative factor of the upper bound.
Since both inequalities hold for every $\theta$, whereas $\bar{P}_{e,u}(R,n)$
is independent of $\theta$, we have actually
proved the inequalities
\begin{equation}
\bar{P}_{e,u}(R,n)\le 2\cdot\min_{\theta\in\Theta}\bar{P}_{e,\theta}(R+\Delta_n,n)
\end{equation}
and 
\begin{equation}
\bar{P}_{e,u}(R,n)\le 2\cdot 2^{n\Delta_n}\cdot\min_{\theta\in\Theta}\bar{P}_{e,\theta}(R,n).
\end{equation}
This completes the proof of Theorem 1. $\Box$

One of the elegant points in \cite{Ziv85} is that the universality 
of the proposed decoding metric, in the random coding error exponent
sense, is proved using a comparative analysis, without recourse to an explicit
derivation of the random coding
error exponent of the optimum decoder. The above proof of Theorem 1 has the
same feature. However, thanks to Shulman's lower
bound on the probability of a union of events, the proof here is both simpler and more
general than in \cite{Ziv85}, in several respects: (i) it allows a general
random coding distribution $Q$, not just the uniform distribution, 
(ii) it requires only pairwise independence
and not full independence between the codewords, and (iii) it assumes nothing
concerning the underlying channel. Indeed, it will be seen shortly How Ziv's
universal decoding metric is obtained as a special case of our approach.

We summarize a few important points:
\begin{enumerate}
\item We have defined a fairly general framework for universal decoding, allowing a
general random coding distribution $Q$, a general channel, and a
a general family of decoding
metrics $\{m_\theta,~\theta\in\Theta\}$. 
Most of the previous works in universal decoding, mentioned in the
second and the third paragraphs of the Introduction, relate to the special case
where the ML decoder for the given channel is equivalent to $m_\theta$
for a certain choice of $\theta$.
\item Another special case that falls within our framework is mismatched
decoding: In this case, $\Theta$ is a singleton and the unique decoding metric $m_\theta$ 
in this singleton is different
from the ML decoding metric of the actual channel.
\item Yet another special case is the case where the channel is deterministic.
This is partially related to the ``individual channel'' paradigm due to
Lomnitz and Feder (see, e.g., \cite{LF12a}, \cite{LF12b} among many other
papers), Misra and Weissman \cite{MW12}, and Shayevitz and Feder \cite{SF05}.
The main difference is that here, we are not concerned with universality of the
encoder, as we simply assume a fixed random coding distribution. In the absence of
feedback, there is no hope for universal encoding.
\end{enumerate}

\section{Useful Approximations of the Universal Decoding Metric}

In some situations, it may not be a trivial task to evaluate
$Q[\calT(\bx|\by)]$, which is needed in order to implement the
proposed universal decoding metric.
Suppose, however, that
one can uniformly lower bound $Q[\calT(\bx|\by)]=\exp\{-nU(\bx,\by)\}$ by
$\exp\{-nU'(\bx,\by)\}$, for some function $U'(\bx,\by)$
which is computable                
and suppose that $U'(\cdot,\cdot)$ is not too large in the sense that it satisfies
the following condition:
\begin{equation}
\label{kraft}
\max_{\by\in\calY^n}\sum_{\bx\in\calX^n}Q(\bx)2^{nU'(\bx,\by)}\le 2^{n\Delta_n'}
\end{equation}
where $\Delta_n'\to 0$.
We argue that in such a case, $U'(\cdot,\cdot)$ can replace
$U(\cdot,\cdot)$ as a universal decoding metric and Theorem 1
remains valid. 

To see why this is true, first observe
that $\bar{\Pi}_{e,\theta}(\bx,\by)$ is trivially
lower bounded by $\exp\{-nU'(\bx,\by)\}$, following (\ref{lb}) and
the very definition of $U'(\bx,\by)$ as an upper bound on $U(\bx,\by)$. As for the upper bound, we have
\begin{eqnarray}
\label{upper1}
\bar{\Pi}_{e,u'}(\bx,\by)&\dfn&\sum_{\{\bx^\prime:~U'(\bx^\prime,\by)\ge
U'(\bx,\by)\}}Q(\bx^\prime)\\
&=&\exp[-nU'(\bx,\by)]\cdot\sum_{\{\bx^\prime:~U'(\bx^\prime,\by)\ge
U'(\bx,\by)\}}Q(\bx^\prime)\exp[nU'(\bx,\by)]\\
&\le&\exp[-nU'(\bx,\by)]\cdot\sum_{\{\bx^\prime:~U'(\bx^\prime,\by)\ge
U'(\bx,\by)\}}Q(\bx^\prime)\exp[nU'(\bx^\prime,\by)]\\
&\le&\exp[-nU'(\bx,\by)]\cdot\sum_{\bx^\prime\in\calX^n}
Q(\bx^\prime)\exp[nU'(\bx^\prime,\by)]\\
&\le&\exp\{-n[U'(\bx,\by)-\Delta_n^\prime]\}.
\end{eqnarray}
Now, the corresponding upper bounds on $\bar{P}_{e,u'}(R,n)$,
in terms of $\min_\theta P_{e,\theta}(R,n)$, are derived
as before, just with $U$ replaced by $U'$. The price of passing from $U$ to $U'$ might be in a slowdown of
the convergence of $\Delta_n^\prime$ vs.\ $\Delta_n$. For example,
$U'$ might correspond to more refined equivalence classes $\{\calT(\bx|\by)\}$.

As an example of the usefulness of this result, let us refer 
to Ziv's universal decoding metric for finite--state channels \cite{Ziv85}. 
In particular, let $\calM$ be
the class of decoding metrics
corresponding to finite--state channels, defined as follows: For a given
$\bx\in\calX^n$ and $\by\in\calY^n$, let $\bs=(s_1,\ldots,s_n)\in\calS^n$
($\calS$ being a finite set), be a sequence generated recursively according to
\begin{equation}
\label{ns}
s_{i+1}=g(x_i,y_i,s_i), ~~~~~i=1,\ldots,n-1, 
\end{equation}
where $s_1$ is some fixed initial
state and $g:\calX\times\calY\times\calS\to\calS$ is a certain next--state
function. Now define
\begin{equation}
m_\theta(\bx,\by)=\sum_{i=1}^n \theta(x_i,y_i,s_i),
\end{equation}
where $\{\theta(x,y,s),~x\in\calX,~y\in\calY,~s\in\calS\}$ are arbitrary real
valued parameters. Similarly as in \cite{Ziv85}, suppose that $Q(\bx)$ is the
uniform distribution over $\calX^n$. Then $Q[\calT(\bx|\by)]$ is proportional
to $|\calT(\bx|\by)|$, but the problem is that here, unlike in Example 1, 
there is no apparent
single--letter expression\footnote{In a nutshell, had there been such a single--letter
expression, one could have easily derived a single--letter expression for the
entropy rate of a hidden Markov process \cite[Section 4.5]{CT06} using the method of types.}
for the exponential growth rate of $|\calT(\bx|\by)|$ in general (unless the
state variable in eq.\ (\ref{ns}) depends solely on the previous state and the
previous channel output). Moreover, $|\calT(\bx|\by)|$ depends on the
next--state function $g$ in eq.\ (\ref{ns}), which is assumed unknown.
Fortunately enough, however,
$|\calT(\bx|\by)|$, in this case, can be lower bounded
\cite[Lemma 1]{Ziv85} by 
\begin{equation}
|\calT(\bx|\by)|\ge 2^{LZ(\bx|\by)-no(n)}, 
\end{equation}
where $LZ(\bx|\by)$ denotes the length (in bits) of the conditional Lempel--Ziv
code (see \cite[proof of Lemma 2]{Ziv85}, \cite{Merhav00}) of $\bx$ when $\by$
is given
as side information at both encoder and decoder.
Consequently, one can upper bound $U(\bx,\by)$ by
\begin{equation}
U'(\bx,\by)=\log|\calX|-\frac{LZ(\bx|\by)}{n}+o(n)
\end{equation}
as our decoding
metric. Indeed, eq.\ (\ref{kraft}) is satisfied by this choice of $U'$ since
\begin{eqnarray}
\sum_{\bx}Q(\bx)2^{nU'(\bx,\by)}&=&
\sum_{\bx}\frac{2^{nU'(\bx,\by)}}{|\calX|^n}\\
&=&\sum_{\bx}2^{-LZ(\bx|\by)+no(n)}\\
&\le& 2^{no(n)},
\end{eqnarray}
where the last equality is Kraft's inequality which holds since
$LZ(\bx|\by)$ is a length function of $\bx$ for every $\by$.
This explains why Ziv's decoder, which selects the message $i$ with
the minimum of $LZ(\bx_i|\by)$, is universally asymptotically optimum in the
random coding exponent sense.
Note that the assumption that $Q$ is uniform is not really essential here.
In fact, $Q$ can also be any exchangeable probability distribution
(i.e., $\bx'$ is a permutation of $\bx$ implies $Q(\bx')=Q(\bx)$). Moreover,
if the state variable $s_i$ includes a component, say, $\sigma_i$, that
is fed merely by $\{x_i\}$ (but not $\{y_i\}$), 
then it is enough that $Q$ would be invariant within
conditional types of $\bx$ given $\bsigma=(\sigma_1,\ldots,\sigma_n)$. 
In such a case, we would have
\begin{equation}
U'(\bx.\by)=-\frac{1}{n}[\log Q(\bx)+LZ(\bx|\by)].
\end{equation}

\section{Extensions}

We now demonstrate how our
method extends to more involved scenarios of communication systems. The first extension
corresponds to random coding distributions that allow access to noiseless
feedback. While this extension is not complicated, it is important from the
operational point of view, because feedback allows the encoder to learn the
channel and thereby to
adapt the random coding distribution
to the channel statistical characteristics. 

Our second extension is
to the problem of universal decoding for multiple access channels (MAC's)
with respect to a given class of decoding metrics (again, without feedback,
but the extension that combines feedback is again straightforward). This extension is
deliberately not provided in full
generality in the sense that we make a certain facilitating assumption on the
structure of
the class of decoding metrics, in order to make the analysis simpler. The main
point here is not the quest for full
generality, but to demonstrate that this extension, even under this
facilitating assumption, is not a trivial task since the
universal decoding metric has to confront three different types of error
events (in the
case of a MAC with two senders): (i) the event were both messages are decoded
incorrectly, (ii) the event where only the message of sender no.\ 1 is 
decoded incorrectly, and (iii) the event where only the message of sender no.\
2 is decoded incorrectly.
As a consequence, it turns out that
the resulting universal decoding metric
is surprisingly different from those of earlier works on universal decoding
for the MAC \cite{LH96}, \cite[Section VIII]{FL98}, \cite{PW85}, mostly
because the
problem setting here is different (and more general) from those of these earlier works (in the
sense that the
universality here is relative to a given class of decoders while the
underlying channel is arbitrary, and not relative to a
given class of channels). While we are not arguing that all the universal
decoders of these previous articles are necessarily suboptimum in our scenario, we are able
to prove the universality only for our own universal decoding metric.

\subsection{Feedback}

In the paradigm of random coding in the presence of feedback, it
is convenient to think of an independent 
random selection of symbols of $\calX$ along
a tree whose branches are labeled by 
$$\{y_1\}, \{y_1,y_2\},\ldots,
\{y_1,\ldots,y_{n-1}\},$$ 
for all possible outcomes of these vectors.
Accordingly, the random coding distribution $Q(\bx)$ is replaced by
\begin{equation}
Q(\bx|\by)\dfn\prod_{i=1}^nQ(x_i|x^{i-1},y^{i-1}).
\end{equation}
Thus, each message $i\in\{1,2,\ldots,M\}$ is represented by a complete tree
of depth $n$ and $|\calY|^{n-1}$ leaves. Theorem 1 and its proof
remain intact with $Q(\cdot)$ being replaced by $Q(\cdot|\by)$ in all places.
Thus, the universal decoding metric is redefined as 
\begin{equation}
U(\bx,\by)=
-\frac{1}{n}\log Q[\calT(\bx|\by)|\by], 
\end{equation}
the expectation in eqs.\ (\ref{ub}) and (\ref{lb})
is redefined w.r.t.\
\begin{equation}
P(\bx,\by)=\prod_{i=1}^n [Q(x_i|x^{i-1},y^{i-1})P(y_i|x^i,y^{i-1})], 
\end{equation} 
and in condition (\ref{kraft}), $Q(\bx)$ is replaced by $Q(\bx|\by)$.

One might limit the structure of the feedback, for example, by letting
each $Q(\cdot|x^{i-1},y^{i-1})$ depend on 
$(x^{i-1},y^{i-1})$ only via a state variable $t_i$ fed by these
two sequences, i.e., 
\begin{equation}
t_i=g(t_{i-1},x_{i-1},y_{i-1}), 
\end{equation}
that is
\begin{equation}
\label{qt}
Q(\bx|\by)=\prod_{i=1}^nQ(x_i|x^{i-1},y^{i-1})=\prod_{i=1}^nQ(x_i|t_i).
\end{equation}
In the above example of decoding metrics 
corresponding to finite--state channels, one can refine the equivalence
classes to include the information about $t_i$ (see Section 4), and then $Q$
would be invariant within
a type class $T_{\bx|\by,\bs,\bt}$, where $\bt=(t_1,\ldots,t_n)$. In this case, the decoding metric
$U'$ would become 
\begin{equation}
U'(\bx,\by)=-\frac{1}{n}[\log Q(\bx|\by)+LZ(\bx|\by)],
\end{equation}
where $Q(\bx|\by)$ is understood to be defined according to eq.\ (\ref{qt}).

\subsection{The Multiple Access Channel}

Consider an arbitrary multiple access channel (MAC), namely, a
channel with two inputs, $\bx_1$ and $\bx_2$, and
one output $\by$. The two inputs are used by two different users which do not
cooperate. User no.\ 1 generates $M_1=2^{nR_1}$ independent codewords,
$\bx_1(1),\ldots,\bx_1(M_1)$, using a
random coding distribution $Q_1$, and
user no.\ 2 generates $M_2=2^{nR_2}$ independent
codewords, $\bx_2(1),\ldots,\bx_2(M_2)$, using a
random coding distribution $Q_2$.\footnote{We should point out that a more general
model definition should allow time--sharing, which means that the codewords of
both users should be drawn conditionally independently given a sequence $\bs$
the designates the time--sharing protocol known to all parties.
This will just amount to conditioning many quantities on $\bs$. 
For the sake of
simplicity of the exposition, we will not add this conditioning on $\bs$.}

We define a class $\calM$ of decoding metrics
$\{m_\theta(\bx_1,\bx_2,\by),~\theta\in\Theta\}$. Decoder $\calD_\theta$
picks the pair of messages $(\bx_1(i),\bx_2(j))$,
$i\in\{1,\ldots,M_1\}$,
$j\in\{1,\ldots,M_2\}$, which
maximizes $m_\theta(\bx_1(i),\bx_2(j),\by)$. We assume that the random coding
ensemble and the class of decoders is such that for every $\by$,
$m_\theta(\bX_1(i),\bX_2(j),\by)$ and
$m_\theta(\bX_1(i'),\bX_2(j'),\by)$ are statistically independent whenever
$(i,j)\neq (i',j')$. While this requirement is easily satisfied when the both $i\ne i'$ and
$j\ne j'$ (for example, when all codewords are drawn by independent random
selection), it is less obvious for combinations of pairs $(i,j)$ and $(i',j')$
for which either $i=i'$ or $j=j'$ (but, of course, not both). Still, this
requirement is satisfied, for example, if
$\calX_1=\calX_2=\{0,1,\ldots,K-1\}$
(or the continuous interval $[0,A]$),
$Q_1$ and $Q_2$ are both uniform across the alphabet, and
$m_\theta(\bx_1,\bx_2,\by)$ depends
on $\bx_1$ and $\bx_2$ only via $\bx_1\oplus\bx_2$, where $\oplus$ denotes
addition modulo $K$ (or addition modulo $A$, in the example of the continuous case). 
Decoding metrics with this property are motivated by classes of multiple
access channels,
$P(\by|\bx_1,\bx_2)$, in which the users interfere with each other additively,
i.e., $P(\by|\bx_1,\bx_2)=W(\by|\bx_1\oplus\bx_2)$.
Still, the dependence of $\by$ on $\bx_1\oplus\bx_2$ can be arbitrary.
In other words, the channel is known to depend only on the
modulo 2 sum of the inputs, but the form of this dependence may not be known.
Another example where the above independence requirement is met
is when $\calX_1=\calX_2=\{-1,+1\}$ and $m_\theta$ depends on $\bx_1$ and
$\bx_2$ only via their component-wise product $\bx_1\cdot\bx_2$.

We now define three kinds of equivalence classes:
\begin{eqnarray}
\calT(\bx_1,\bx_2|\by)&=&\left\{(\bx_1',\bx_2'):~\forall\theta\in\Theta~m_\theta(\bx_1',\bx_2',\by)=
m_\theta(\bx_1,\bx_2,\by)\right\}\\
\calT(\bx_1|\bx_2,\by)&=&\left\{\bx_1':~\forall\theta\in\Theta~m_\theta(\bx_1',\bx_2,\by)
=m_\theta(\bx_1,\bx_2,\by)\right\}\nonumber\\
&=&\{\bx_1':~(\bx_1',\bx_2)\in\calT(\bx_1,\bx_2|\by)\}\\
\calT(\bx_2|\bx_1,\by)&=&\left\{\bx_2':~\forall\theta\in\Theta~m_\theta(\bx_1,\bx_2',\by)
=m_\theta(\bx_1,\bx_2,\by)\right\}\nonumber\\
&=&\{\bx_2':~(\bx_1,\bx_2')\in\calT(\bx_1,\bx_2|\by)\}.
\end{eqnarray}
We also assume, as before, that for every $\by$, the number of different
type classes $\{\calT(\bx_1,\bx_2|\by)\}$ is upper bounded by
$2^{n\Delta_n}$.
Next, define the following functions:
\begin{eqnarray}
U_0(\bx_1,\bx_2,\by)&=&-\frac{1}{n}\log\left\{(Q_1\times
Q_2)[\calT(\bx_1,\bx_2|\by)]\right\}\\
U_1(\bx_1,\bx_2,\by)&=&-\frac{1}{n}\log Q_1
[\calT(\bx_1|\bx_2,\by)]\\
U_2(\bx_1,\bx_2,\by)&=&-\frac{1}{n}\log Q_2
[\calT(\bx_2|\bx_1,\by)].
\end{eqnarray}
What makes the MAC interesting, in the context of universal decoding, is that
the universal decoder has to cope with three different types of errors:
(i) both messages are decoded incorrectly,
(ii) the message of user no.\ 2 is decoded correctly, but that of user no.\
1 is not, and (iii) like (ii), but with the roles of the users swapped.
From Theorem 1 and its proof (after an obvious modification), it is apparent that
had only errors of type (i)
existed, then $U_0$ could have been a universal decoding metric. Similarly,
had only errors of type (ii) existed, then $U_1$ could be a universal decoding metric,
and by the same token, for error of type (iii) alone, one would use $U_2$. 
However, in reality, all three types of error events might occur and we need one universal
decoding metric that handles all of them at the same time. The question is
then how to combine $U_0$, $U_1$ and $U_2$ into one metric that would work at
least as well as the best decoder in the given class.

The answer turns out to be the following: Define the universal decoding metric
as
\begin{equation}
U(\bx_1,\bx_2,\by)=\min\left\{[U_0(\bx_1,\bx_2,\by)-R_1-R_2],[U_1(\bx_1,\bx_2,\by)-R_1],
[U_2(\bx_1,\bx_2,\by)-R_2]\right\}.
\end{equation}
We argue that $U(\bx_1,\bx_2,\by)$ competes favorably with the best $m_\theta$ in a sense
analogous to that asserted in Theorem 1. This decoding metric is different
from the universal decoding metrics used for the MAC, for example, in \cite{PW85} and
\cite{LH96}, which were based on the MMI decoder and the minimum empirical conditional entropy
(minimum equivocation) rule, respectively. It is not argued here that these
decoding rules are necessarily suboptimal in the present setting, 
but on the other hand, we do not have a
proof that they compete favorably with the best decoder in the class $\calM$.
The remaining part of this section is devoted to a description of
the main modifications and extensions needed in the proof of Theorem 1 in
order to prove the universality of $U(\bx_1,\bx_2,\by)$ for the MAC.

The pairwise probability of type (i) error
for an arbitrary decoder in the reference class $\calM$
is lower bounded by
\begin{eqnarray}
P_{e,\theta}^{(i)}(\bx_1,\bx_2,\by)&\dfn&\sum_{\{\bx_1',\bx_2':~m_\theta(\bx_1',\bx_2',\by)\ge
m_\theta(\bx_1,\bx_2,\by)\}}Q_1(\bx_1')Q_2(\bx_2')\\
&\ge&\sum_{(\bx_1',\bx_2')\in\calT(\bx_1,\bx_2|\by)}Q_1(\bx_1')Q_2(\bx_2')\\
&=&(Q_1\times Q_2)[\calT(\bx_1,\bx_2|\by)]\\
&=&2^{-nU_0(\bx_1,\bx_2,\by)}.
\end{eqnarray}
As for the pairwise error probability of type (ii),
we have
\begin{eqnarray}
P_{e,\theta}^{(ii)}(\bx_1,\bx_2,\by)&\dfn&\sum_{\{\bx_1':~m_\theta(\bx_1',\bx_2,\by)\ge
m_\theta(\bx_1,\bx_2,\by)\}}Q_1(\bx_1')\\
&\ge&\sum_{\bx_1'\in\calT(\bx_1|\bx_2,\by)}Q_1(\bx_1')\\
&=&Q_1[\calT(\bx_1|\bx_2,\by)]\\
&=&2^{-nU_1(\bx_1,\bx_2,\by)}.
\end{eqnarray}
and similarly, for type (iii):
\begin{equation}
P_{e,\theta}^{(iii)}(\bx_1,\bx_2,\by)\geq 2^{-nU_2(\bx_1,\bx_2,\by)}.
\end{equation}
Let $\calI$ be half of the set $\{1,2,\ldots,M_1\}-\{i\}$ and let
$\calJ$ be half of the set $\{1,2,\ldots,M_2\}-\{j\}$ ($i$ and $j$ being the
correct messages of the two senders).
Let $\calA$ be the set of all $(M_1-1)(M_2-1)/4$ pairs 
$(i',j')\in\calI^c\times\calJ^c$, where both $i'\ne i$ and $j'\ne j$.
Under our above assumptions, the following is true: given $(\bx_1(i),\bx_2(j),\by)$, the events
$$\{m_\theta(\bX_1(i'),\bx_2(j),\by)\ge m_\theta(\bx_1(i),\bx_2(j),\by)\}_
{i'\in\calI},$$
$$\{m_\theta(\bx_1(i),\bX_2(j'),\by)\ge m_\theta(\bx_1(i),\bx_2(j),\by)\}_
{j'\in\calJ},$$ 
and
$$\{m_\theta(\bX_1(i'),\bX_2(j'),\by)\ge m_\theta(\bx_1(i),\bx_2(j),\by)\}_
{(i',j')\in\calA}$$ 
are all pairwise independent. Defining the set of pairs
$\calB=\calA\cup[\{i\}\times \calJ]\cup[\calI\times\{j\}]$,
the total probability of error,
associated with the decoder $\calD_\theta$, is lower bounded as follows:
\begin{eqnarray}
& &\bar{P}_{e,\theta}(R_1,R_2,n)\\
&\dfn& \frac{1}{M_1M_2}\sum_{i=1}^{M_1}\sum_{j=1}^{M_2}\mbox{Pr}\bigcup_{(i',j')\ne (i,j)}
\left\{m_\theta(\bX_1(i'),\bX_2(j'),\bY)\ge
m_\theta(\bX_1(i),\bX_2(j),\bY)\bigg|(i,j)~\mbox{sent}\right\}\\
&\ge& \frac{1}{M_1M_2}\sum_{i=1}^{M_1}\sum_{j=1}^{M_2}
\mbox{Pr}\bigcup_{(i',j')\in \calB}\left\{
m_\theta(\bX_1(i'),\bX_2(j'),\bY)\ge
m_\theta(\bX_1(i),\bX_2(j),\bY)\bigg|(i,j)~\mbox{sent}\right\}\\
&\gexe&\bE\min\left\{1,\frac{(M_1-1)(M_2-1)}{4}\cdot 2^{-nU_0(\bX_1,\bX_2,\bY)}+\right.\nonumber\\
& &\frac{(M_1-1)}{2}\cdot 2^{-nU_1(\bX_1,\bX_2,\bY)}+
\left.\frac{(M_2-1)}{2}\cdot 2^{-nU_2(\bX_1,\bX_2,\bY)}\right\}\\
&\exe&\bE\min\left\{1,2^{-n[U_0(\bX_1,\bX_2,\bY)-R_1-R_2]}+\right.\nonumber\\
& &\left.2^{-n[U_1(\bX_1,\bX_2,\bY)-R_1]}+
2^{-n[U_2(\bX_1,\bX_2,\bY)-R_2]}\right\}\\
&\exe&\bE\min\left\{1,2^{-nU(\bX_1,\bX_2,\bY)}\right\}\\
&=&\bE\{ 2^{-n[U(\bX_1,\bX_2,\bY)]_+}\},
\end{eqnarray}
where the second inequality is again due 
to Shulman \cite[Lemma A.2]{Shulman03}.
Consider now the function $U(\bx_1,\bx_2,\by)$ as a universal decoding metric.
Then, we have the following:
\begin{eqnarray}
\label{uppermac1}
P_{e,u}^{(i)}(\bx_1,\bx_2,\by)&\dfn&\sum_{\{(\bx_1',\bx_2'):~U(\bx_1',\bx_2',\by)\ge
U(\bx_1,\bx_2,\by)\}}Q_1(\bx_1')Q_2(\bx_2')\\
&=&\sum_{\{\calT(\bx_1',\bx_2'|\by):~U(\bx_1',\bx_2',\by)\ge U(\bx_1,\bx_2,\by)\}}
\sum_{(\tilde{\bx}_1,\tilde{\bx}_2)\in\calT(\bx_1',\bx_2'|\by)}Q_1(\tilde{\bx}_1)Q_2(\tilde{\bx}_2)\\
&=&\sum_{\{\calT(\bx_1',\bx_2'|\by):~U(\bx_1',\bx_2',\by)\ge U(\bx_1,\bx_2,\by)\}}
(Q_1\times Q_2)[\calT(\bx_1',\bx_2'|\by)]\\
&\le&\sum_{\{\calT(\bx_1',\bx_2'|\by):~U(\bx_1',\bx_2',\by)\ge U(\bx_1,\bx_2,\by)\}}
\exp[-nU_0(\bx_1',\bx_2',\by)].
\end{eqnarray}
Similarly,
\begin{eqnarray}
\label{uppermac2}
P_{e,u}^{(ii)}(\bx_1,\bx_2,\by)&\dfn&\sum_{\{\bx_1':~U(\bx_1',\bx_2,\by)\ge
U(\bx_1,\bx_2,\by)\}}Q_1(\bx_1')\\
&=&\sum_{\{\calT(\bx_1'|\bx_2,\by):~U(\bx_1',\bx_2,\by)\ge
U(\bx_1,\bx_2,\by)\}}
\sum_{\tilde{\bx}_1\in\calT(\bx_1'|\bx_2,\by)}Q_1(\tilde{\bx}_1)\\
&\le&\sum_{\{\calT(\bx_1',\bx_2'|\by):~U(\bx_1',\bx_2',\by)\ge
U(\bx_1,\bx_2,\by)\}}
\sum_{\tilde{\bx}_1\in\calT(\bx_1'|\bx_2',\by)}Q_1(\tilde{\bx}_1)\\
&=&\sum_{\{\calT(\bx_1',\bx_2'|\by):~U(\bx_1',\bx_2',\by)\ge
U(\bx_1,\bx_2,\by)\}}
Q_1[\calT(\bx_1'|\bx_2',\by)]\\
&\le&\sum_{\{\calT(\bx_1',\bx_2'|\by):~U(\bx_1',\bx_2',\by)\ge
U(\bx_1,\bx_2,\by)\}}
\exp[-nU_1(\bx_1',\bx_2',\by)], 
\end{eqnarray}
and by the same token,
\begin{equation}
P_{e,u}^{(iii)}(\bx_1,\bx_2,\by)\le
\sum_{\{\calT(\bx_1',\bx_2'|\by):~U(\bx_1',\bx_2',\by)\ge
U(\bx_1,\bx_2,\by)\}}
\exp[-nU_2(\bx_1',\bx_2',\by)]. 
\end{equation}
Now,
\begin{eqnarray}
& &\bar{P}_{e,u}(R_1,R_2,n)\\
&=&\frac{1}{M_1M_2}\sum_{i=1}^{M_1}\sum_{j=1}^{M_2}\mbox{Pr}\left\{\bigcup_{(i',j')\ne (i,j)}
U(\bX_1(i'),\bX_2(j'),\bY)\ge
U(\bX_1(i),\bX_2(j),\bY)\bigg|(i,j)~\mbox{sent}\right\}\\
&\exe&\bE\min\left\{1,2^{n(R_1+R_2)}P_{e,u}^{(i)}(\bX_1,\bX_2,\bY)+\right.\nonumber\\
& &\left. 2^{nR_1}P_{e,u}^{(ii)}(\bX_1,\bX_2,\bY)+
2^{nR_2}P_{e,U}^{(iii)}(\bX_1,\bX_2,\bY)\right\}\\
&\exe&\bE\min\left\{1,\sum_{\{\calT(\bx_1',\bx_2'|\bY):~U(\bx_1',\bx_2',\bY)\ge
U(\bX_1,\bX_2,\bY)\}}\left[2^{-n[U_0(\bx_1',\bx_2',\bY)-R_1-R_2]}+\right.\right.\\
& &\left.\left. 2^{-n[U_1(\bx_1',\bx_2',\bY)-R_1]}+
2^{-n[U_2(\bx_1',\bx_2',\bY)-R_2]}\right]\right\}\\
&\exe&\bE\min\left\{1,\sum_{\{\calT(\bx_1',\bx_2'|\bY):~U(\bx_1',\bx_2',\bY)\ge
U(\bX_1,\bX_2,\bY)\}}2^{-nU(\bx_1',\bx_2',\bY)}\right\}\\
&\le&\bE\min\left\{1,\sum_{\{\calT(\bx_1',\bx_2'|\bY):~U(\bx_1',\bx_2',\bY)\ge
U(\bX_1,\bX_2,\bY)\}}2^{-nU(\bX_1,\bX_2,\bY)}\right\}\\
&\le&\bE\min\left\{1,\sum_{\{\calT(\bx_1',\bx_2'|\bY)
\}}2^{-nU(\bX_1,\bX_2,\bY)}\right\}\\
&\le&\bE\min\left\{1,
2^{-n[U(\bX_1,\bX_2,\bY)-\Delta_n]}\right\}\\
&=&\bE\left\{
2^{-n[U(\bX_1,\bX_2,\bY)-\Delta_n]_+}\right\},
\end{eqnarray}
which is of the same exponential order as the lower bound on
$\bar{P}_{e,\theta}(R,n)$, and hence
$\bar{P}_{e,u}(R,n)$ is exponentially at least as small as
$\min_{\theta\in\Theta}\bar{P}_{e,\theta}(R,n)$.

Similarly as in Section 4, suppose that $U_0$, $U_1$ and $U_2$ can be uniformly
upper bounded by $U_0'$, $U_1'$ and $U_2'$, respectively, and assume that:
\begin{eqnarray}
\max_{\by}\sum_{\bx_1,\bx_2}Q_1(\bx_1)Q_2(\bx_2)2^{nU_0'(\bx_1,\bx_2,\by)}&\lexe&1\\
\max_{\bx_2,\by}\sum_{\bx_1}Q_1(\bx_1)2^{nU_1'(\bx_1,\bx_2,\by)}&\lexe&1\\
\max_{\bx_1,\by}\sum_{\bx_2}Q_2(\bx_1)2^{nU_2'(\bx_1,\bx_2,\by)}&\lexe&1.
\end{eqnarray}
Then, $U_0'$, $U_1'$ and $U_2'$ can replace $U_0$, $U_1$ and $U_2$,
respectively, in the universal decoding metric, denoted in turn by $U'$,
and the upper and lower bounds continue to hold with $U'$ replacing $U$.
The lower bounds on $P_{e,\theta}^{(i)}(\bx_1,\bx_2,\by)$,
$P_{e,\theta}^{(ii)}(\bx_1,\bx_2,\by)$, and
$P_{e,\theta}^{(iii)}(\bx_1,\bx_2,\by)$, in terms of $U_0'$, $U_1'$ and
$U_2'$, respectively,
are trivial, of course. As for the upper bounds on
$P_{e,u'}^{(i)}(\bx_1,\bx_2,\by)$,
$P_{e,u'}^{(ii)}(\bx_1,\bx_2,\by)$, and
$P_{e,u'}^{(iii)}(\bx_1,\bx_2,\by)$, we proceed similarly as follows:
\begin{eqnarray}
\label{uppermac1p}
P_{e,u'}^{(i)}(\bx_1,\bx_2,\by)&=&\sum_{\{(\bx_1',\bx_2'):~U'(\bx_1',\bx_2',\by)\ge
U'(\bx_1,\bx_2,\by)\}}Q_1(\bx_1')Q_2(\bx_2')\\
&=&\sum_{\{\calT(\bx_1',\bx_2'|\by):~U'(\bx_1',\bx_2',\by)\ge
U'(\bx_1,\bx_2,\by)\}}
\sum_{(\tilde{\bx}_1,\tilde{\bx}_2)\in\calT(\bx_1',\bx_2'|\by)}Q_1(\tilde{\bx}_1)Q_2(\tilde{\bx}_2)\\
&=&\sum_{\{\calT(\bx_1',\bx_2'|\by):~U'(\bx_1',\bx_2',\by)\ge
U'(\bx_1,\bx_2,\by)\}}2^{-nU_0'(\bx_1',\bx_2',\by)}\times\nonumber\\
& &\sum_{(\tilde{\bx}_1,\tilde{\bx}_2)\in\calT(\bx_1',\bx_2'|\by)}Q_1(\tilde{\bx}_1)Q_2(\tilde{\bx}_2)
2^{nU_0'(\bx_1',\bx_2',\by)}\\
&=&\sum_{\{\calT(\bx_1',\bx_2'|\by):~U'(\bx_1',\bx_2',\by)\ge
U'(\bx_1,\bx_2,\by)\}}2^{-nU_0'(\bx_1',\bx_2',\by)}\times\nonumber\\
& &\sum_{(\tilde{\bx}_1,\tilde{\bx}_2)\in\calT(\bx_1',\bx_2'|\by)}Q_1(\tilde{\bx}_1)Q_2(\tilde{\bx}_2)
2^{nU_0'(\tilde{\bx}_1,\tilde{\bx}_2,\by)}\\
&\lexe&\sum_{\{\calT(\bx_1',\bx_2'|\by):~U'(\bx_1',\bx_2',\by)\ge
U'(\bx_1,\bx_2,\by)\}}2^{-nU_0'(\bx_1',\bx_2',\by)}
\end{eqnarray}
and similar treatments hold for $P_{e,u'}^{(ii)}(\bx_1,\bx_2,\by)$, and
$P_{e,u'}^{(iii)}(\bx_1,\bx_2,\by)$. This suggests that, in the case
where the class $\calM$ is based on finite--state machines, 
\begin{equation}
m_\theta(\bx_1,\bx_2,\by)=\sum_{i=1}^n\theta(x_{1,i},x_{2,i},y_i,s_i),~~~~~~~
s_{i+1}=g(x_{1,i},x_{2,i},y_i,s_i)
\end{equation}
and $Q_1$ and $Q_2$ are uniform distributions within single type classes,
one may use
$LZ(\bx_1,\bx_2|\by)$, 
$LZ(\bx_1|\bx_2,\by)$ and
$LZ(\bx_2|\bx_1,\by)$ in the relevant places, i.e., the universal decoding
metric would be
\begin{eqnarray}
U'(\bx_1,\bx_2,\by)&=&\min\left\{\left[\hat{H}_{\bx_1}(X)+\hat{H}_{\bx_2}(X)-\frac{
LZ(\bx_1,\bx_2|\by)}{n}-R_1-R_2\right],
\right.\nonumber\\
& &\left[\hat{H}_{\bx_1}(X)-\frac{
LZ(\bx_1|\bx_2,\by)}{n}-R_1\right],\nonumber\\
& &\left.\left[\hat{H}_{\bx_2}(X)-\frac{LZ(\bx_2|\bx_1,\by)}{n}-R_2\right]\right\}.
\end{eqnarray}
Thus, we observe that our approach suggests a systematic method to extend
earlier results to more involved scenarios, like that of the MAC.

\section*{Acknowledgment}

Interesting discussions with Meir Feder and Yuval Lomnitz are acknowledged
with thanks.


\end{document}